# Bringing coherent closer to compute: coherent-lite multidimensional on-chip interconnect beyond 100 terabits per second

Yongzhu Hu[1,4], Aolong Sun[1,2,4], Huayuan Qin[1], An Yan[1], Jialin He[1], Yuqin Yuan[1], Junhao Zhao[1], Chao Shen[1], Ziwei Li[1], Peng Sun[3], Hui Chen[3], Chao Li[3], Xu Zhang[3], Bowei Dong[2], Zhixue He[3*], Nan Chi[1], Shaohua Yu[3], Junwen Zhang[1*]

[1] College of Future Information Technology, Fudan University, Shanghai, China

[2] Institute of Microelectronics, Agency for Science, Technology and Research, Singapore

[3] Peng Cheng Laboratory, Shenzhen, China

[4] These authors contributed equally to this work: Yongzhu Hu, Aolong Sun

[*] Corresponding authors: junwenzhang@fudan.edu.cn; hezhx01@pcl.ac.cn

## Abstract

The continued scaling of artificial intelligence (AI) demands optical interconnects that deliver higher bandwidth density within stringent electrical-bandwidth and energy budgets. Coherent optics offers a route beyond the per-lane bandwidth constraints of intensity-modulation direct-detection links, yet its receiver complexity and power overhead have largely confined its use to longer reaches. Here we bring coherent optics closer to compute by demonstrating a coherent-lite multidimensional on-chip interconnect that combines wavelength- and mode-division multiplexing with self-homodyne detection. The architecture co-transmits the local oscillator (LO) and parallel data channels within a single silicon waveguide, using polarization-division routing to preserve LO purity and support high-order modulation. Based on this principle, we develop an inverse-designed dual-polarization mode multiplexer that provides over 37 dB inter-polarization isolation across the C band in an $8 \times 6$ $\mu m^2$ footprint, enabling millimetre-scale on-chip multimode routing. Without power-intensive digital carrier recovery, the architecture delivers a single-wavelength line rate of 2 Tb $s^{-1}$ and scales to an aggregate line rate of 103.96 Tb $s^{-1}$ across four modes and 100 wavelengths, using formats up to 64-ary quadrature amplitude modulation (QAM). These results extend on-chip coherent interconnects into the hundred-terabit-per-second regime, offering a scalable route to bandwidth-dense optical connectivity for next-generation computing systems.

## Introduction

The rapid expansion of artificial intelligence (AI) workloads is making interconnect bandwidth density and energy efficiency central to the scalability of computing systems[1,2]. These systems depend on an interconnect hierarchy spanning data centres, racks and accelerator packages (Fig. 1a). Frequency-dependent attenuation and routing constraints increasingly limit the capacity of copper-based electrical links, driving optical connectivity closer to compute[3,4]. Silicon photonics supports this transition by combining optical bandwidth and multiplexing with compact devices compatible with complementary metal–oxide–semiconductor (CMOS) fabrication[5,6], enabling co-packaged optics, optical I/O and photonic interposers[7]. For these short-reach optical links, intensity modulation and direct detection (IM-DD) offers a simple and cost-effective transceiver architecture[8]. As optical interfaces become more densely integrated, capacity per fibre or waveguide becomes a key determinant of aggregate bandwidth within a constrained routing footprint[9,10]. With per-lane symbol rates approaching 200 Gbaud, however, further capacity scaling through intensity-only modulation places ever-higher demands on electrical and optoelectronic bandwidth[11]. Continued progress therefore calls for higher spectral efficiency and greater capacity per optical path within stringent energy and complexity budgets.

Coherent optics addresses the bandwidth-efficiency requirement by exploiting both the amplitude and phase of the optical field, enabling high-order quadrature amplitude modulation (QAM) to increase the information carried per symbol[12]. Its capacity-scaling potential has been demonstrated in long-haul optical links achieving aggregate capacities of hundreds of terabits per second[13,14]. However, conventional full-coherent architectures typically rely on an independent local oscillator (LO) laser and digital carrier recovery to compensate for transmitter–LO frequency offset and relative phase noise (Fig. 1b)[15]. The resulting hardware complexity and power overhead are difficult to reconcile with the energy budgets of short-reach interconnects, motivating coherent-lite architectures[16,17]. Self-homodyne transmission provides one such route by delivering a transmitter-derived optical reference that serves as the LO at the receiver, eliminating the need for an independent LO laser and enabling reception without power-intensive digital carrier recovery[18–20]. Demonstrations combining high-order modulation with wavelength-space parallelism in few-mode fibre links highlight its potential for high-capacity scale-out optical interconnects[21].

Achieving high bandwidth density in on-chip self-homodyne interconnects, however, requires the transmitted LO and parallel data channels to be jointly accommodated within a tightly constrained footprint[22]. On-chip mode-division multiplexing (MDM) offers compact spatial parallelism[23,24], with metamaterial-enabled and inverse-designed mode multiplexers supporting high channel counts and capacities up to tens of terabits per second within a single waveguide[25–27]. Yet architectures developed primarily for data-channel scaling do not inherently ensure the isolation needed to preserve a co-propagating LO. The nonuniform dielectric profiles used for mode conversion can produce nonzero weighted overlaps between otherwise orthogonal modal fields, permitting residual intermodal coupling (Fig. 1c)[28]. Any resulting leakage of modulated signals into the LO generates signal-signal beating interference (SSBI) upon detection (Supplementary Note 2), reducing the effective receiver signal-to-noise ratio (SNR) and limiting the achievable modulation order[29]. A previous on-chip demonstration that carried the LO in $TE_0$ and data in higher-order TE modes was limited to quadrature phase-shift keying (QPSK), with crosstalk-induced LO contamination degrading transmission performance[30]. Existing on-chip interconnects therefore face an unresolved challenge in delivering a high-purity LO alongside dense spatial parallelism, as required to translate coherent spectral efficiency into higher bandwidth density.

Here we demonstrate a coherent-lite multidimensional architecture that unlocks the potential of coherent optics for bandwidth-dense on-chip interconnects by combining wavelength- and mode-division multiplexing with self-homodyne detection (Fig. 1d). The architecture uses polarization-division routing (PDR) to preserve LO purity by assigning orthogonal polarizations to the LO and the spatially multiplexed data. We implement this routing principle with an inverse-designed silicon dual-polarization mode multiplexer (DP-mMUX). Within a footprint of $8 \times 6$ μm$^2$, the device multiplexes a $TM_0$ LO and four $TE_0$–$TE_3$ data modes into a single multimode waveguide. It provides over 37 dB of TM-to-TE isolation across the C band, enabling millimetre-scale on-chip multimode routing of the LO and data. Using this architecture, we demonstrate 100-Gbaud 32-QAM transmission across four spatial modes, delivering a single-wavelength line rate of 2 Tb s$^{-1}$ without digital carrier recovery. We further scale the architecture to 100 wavelength channels with modulation formats up to 64-QAM, achieving an aggregate line rate of 103.96 Tb s$^{-1}$, representing, to our knowledge, the first demonstration of a hundred-terabit-per-second on-chip coherent optical interconnect within a single silicon waveguide. Our work points to a broader role for coherent optics in next-generation interconnects, extending its spectral-efficiency and multidimensional scaling advantages towards increasingly short-reach, bandwidth-dense connectivity.

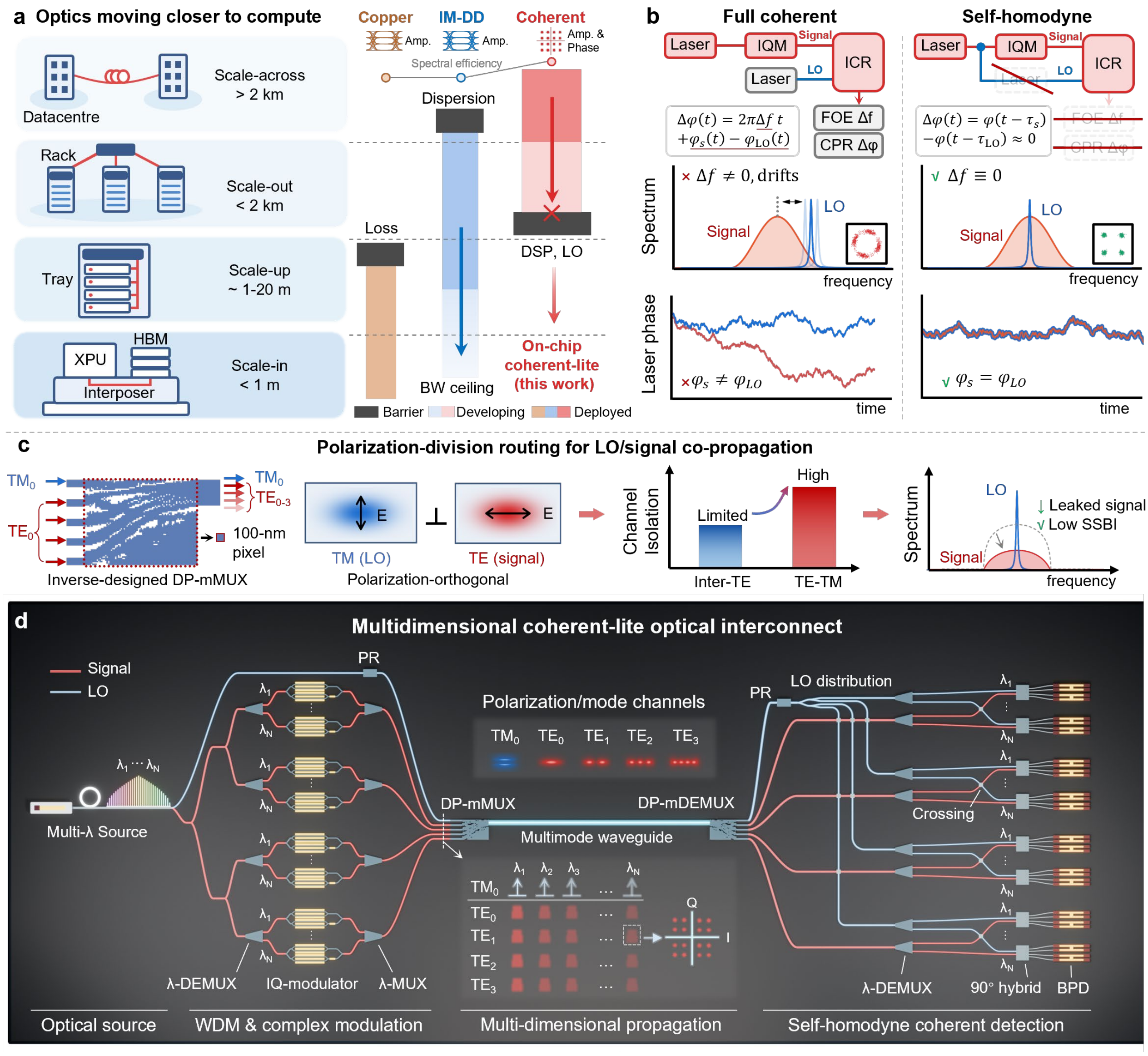

**Fig. 1 Principle and architecture of the multidimensional coherent-lite on-chip interconnect. a,** Current landscape of interconnect technologies across the AI infrastructure hierarchy[1]. Copper links serve scale-in and short scale-up connections, with attenuation limiting longer transmission. IM-DD optics span scale-up and scale-out under dispersion and bandwidth constraints. Coherent optics offers the highest spectral efficiency and capacity potential at scale-out and scale-across. This work extends coherent optics towards scale-in interconnects with an on-chip coherent-lite architecture. XPU, generic processing unit; HBM, high-bandwidth memory. **b,** Full-coherent and self-homodyne detection. Full-coherent receivers require an independent LO laser and digital compensation for transmitter–LO frequency offset and relative phase noise. Self-homodyne detection derives the signal and LO from the same laser, enabling coherent recovery without a separate receiver-side laser, frequency-offset estimation (FOE) or carrier-phase recovery (CPR). IQM, in-phase and quadrature modulator; ICR, integrated coherent receiver. **c,** Polarization-division routing for high-purity LO delivery. An inverse-designed DP-mMUX combines the LO in $TM_0$ and four data channels in $TE_0$–$TE_3$ within a shared multimode waveguide. Polarization orthogonality provides high TE–TM isolation during mode conversion and co-propagation, suppressing modulated-signal leakage into the LO and reducing SSBI upon coherent detection. **d,** Conceptual architecture for a multidimensional coherent-lite on-chip optical interconnect. An integrated optical frequency comb serves as the multi-λ source[9,10], supplying an unmodulated LO and four signal branches. Wavelength channels in each signal branch are separated by a λ-DEMUX, modulated by IQ modulators and recombined by a λ-MUX. The DP-mMUX combines the LO and WDM signals in a single multimode waveguide for multidimensional propagation. PRs provide polarization matching between the TM LO channel and the TE-polarized source and receiver interfaces. At the receiver, the DP-mDEMUX separates the LO and signals, and the LO is distributed to the four receiver branches. The λ-DEMUXs supply corresponding signal–LO wavelength pairs to 90° hybrids and BPDs for self-homodyne coherent detection. λ-MUX/λ-DEMUX, wavelength multiplexer/demultiplexer; PR, polarization rotator; BPD, balanced photodetector.

## Results

### Inverse-designed dual-polarization mode multiplexer

The DP-mMUX is designed to preserve a high-purity $TM_0$ LO while multiplexing four TE data modes ($TE_0$–$TE_3$) into a common multimode bus. It connects five 0.5-μm-wide single-mode inputs to a 1.8-μm-wide bus through an $8 \times 6$ $\mu m^2$ design region (Fig. 2a). In contrast to the residual intermodal coupling that limits same-polarization designs, TE–TM coupling in the DP-mMUX is forbidden by structural symmetry. The device is fabricated in a 220-nm silicon layer between a $SiO_2$ upper cladding and a buried oxide (BOX) layer, with the 100-nm pixel pattern etched through the full silicon thickness (Fig. 2b). This geometry preserves horizontal mirror symmetry, $\sigma_h$, while retaining the in-plane design freedom required for mode conversion. Mirror parity is consequently conserved throughout the patterned region, with the $TE_0$–$TE_3$ and $TM_0$ modes occupying opposite symmetry sectors. Their dielectric-weighted overlap integrand is odd about $\sigma_h$ and integrates to zero for any in-plane pattern (Fig. 2c)[31].

We employ edge-guided analog-and-digital optimization (EG-ADO)[26] to jointly optimize coupling from the five input waveguides to their respective target modes in the multimode bus. This method combines adjoint optimization[32] with edge-guided digitization and restricts binary refinement to edge pixels, reducing the discrete search space while yielding fabrication-compatible pixelated structures (Methods and Supplementary Note 1). Three-dimensional finite-difference time-domain (3D-FDTD) simulations show that the optimized DP-mMUX achieves low insertion loss across all five channels and high inter-polarization isolation over 1500–1600 nm (Fig. 2d), with each input field evolving through the design region into its target bus mode (Fig. 2e). Further analysis over 1530–1570 nm shows that, within commercial foundry tolerances, pixel-size deviations preserve the mirror symmetry and keep TE–TM crosstalk at the numerical floor, whereas sidewall tilt breaks this symmetry and raises the worst-case crosstalk to −36.4 dB. Insertion-loss penalties remain below 0.6 dB in both cases (Supplementary Figs. 2 and 3).

We characterize the fabricated device using a back-to-back MUX–DEMUX pair, with grating-coupler losses removed using reference waveguides (Fig. 2f and Methods). At 1550 nm, the measured single-MUX insertion losses for $TE_0$–$TE_3$ are 0.47, 1.67, 0.51 and 0.79 dB, respectively. At this wavelength, the worst-case intermodal crosstalk in the back-to-back circuit ranges from −20.8 to −9.7 dB across the four TE inputs. Across 1530–1570 nm, the single-MUX insertion loss for $TM_0$ remains below 0.30 dB, and the isolation from the $TM_0$ input to each of the four TE outputs is higher than 37 dB. This exceeds the best-channel isolation of the ultracompact inverse-designed mode multiplexers compared in Fig. 2g[24,26,27,33–37], supporting high-order modulation in self-homodyne transmission.

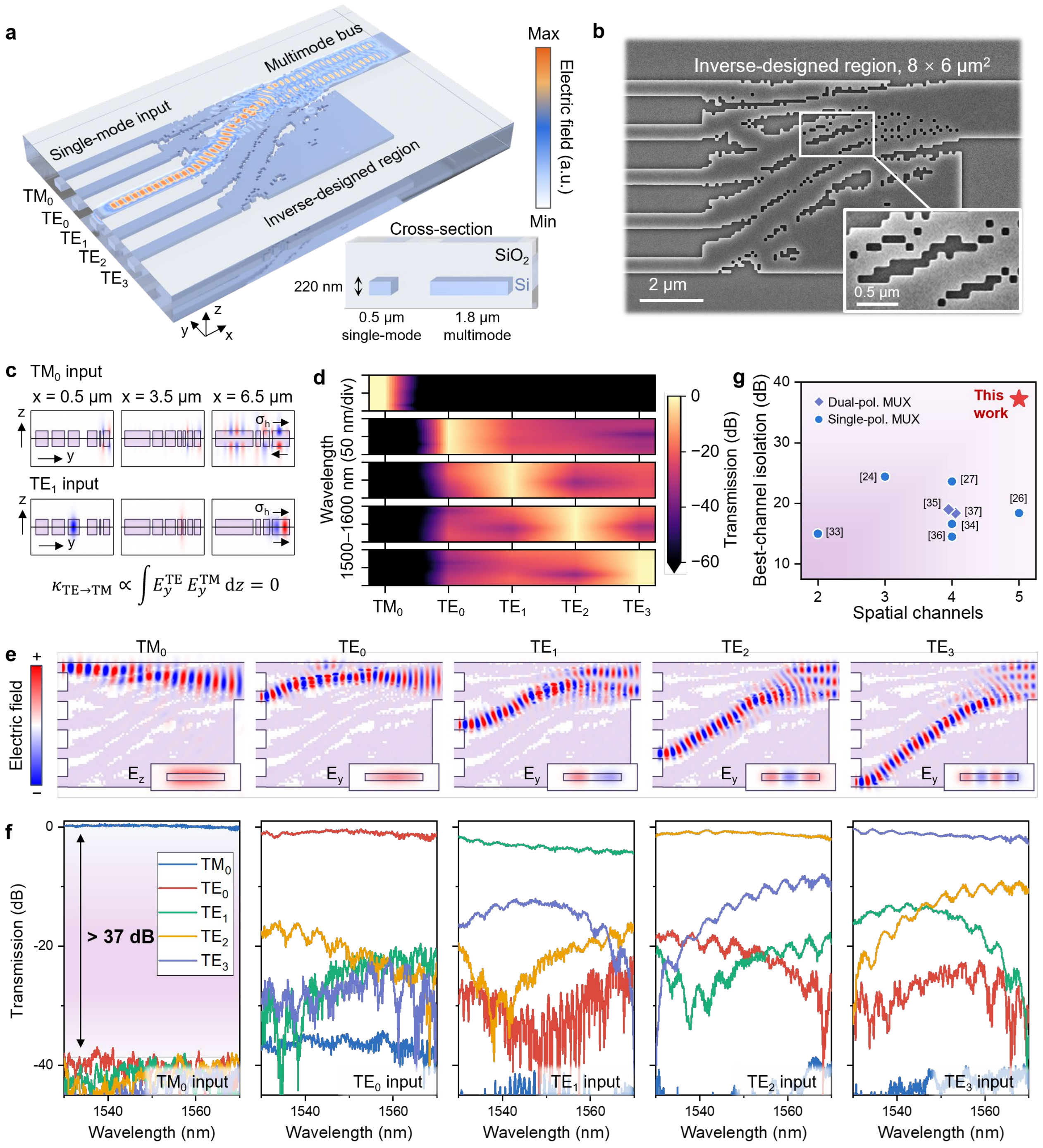

**Fig. 2 Design and characterization of the inverse-designed dual-polarization mode multiplexer. a,** Schematic of the DP-mMUX with a simulated electric-field overlay for the $TE_1$ channel at 1550 nm. Inset, waveguide cross-section. Input ports are labelled by their corresponding target modes in the multimode bus. **b,** Scanning electron micrograph (SEM) of the fabricated DP-mMUX and an enlarged view of the pixel pattern. **c,** Simulated $Re(E_y)$ distributions at 1550 nm in $y$–$z$ planes located 0.5, 3.5 and 6.5 μm from the start of the design region ($x = 0$), showing odd and even parity about $\sigma_h$ for the $TM_0$ (top) and $TE_1$ (bottom) channels, respectively. **d,** Simulated modal transmission matrix over 1500–1600 nm, with rows corresponding to input ports and columns to output modes. **e,** Simulated propagation fields at 1550 nm for the five channels. Insets, output modal profiles. **f,** Measured transmission spectra of the back-to-back circuit for each of the five inputs. Single-MUX insertion losses are estimated as half the corresponding back-to-back losses in dB. **g,** Comparison with reported ultracompact inverse-designed mode multiplexers in terms of spatial channel count and best-channel isolation[24,26,27,33–37]. For each device, we evaluate each mode channel by its minimum isolation across the C band. The highest of these values is reported as the best-channel isolation.

## Polarization-division routing preserves LO purity

We evaluate PDR in self-homodyne transmission using the scheme in Fig. 3a, in which the IQ-modulated signal and the unmodulated LO originate from the same laser and co-propagate through the chip in the TE data modes and the $TM_0$ mode, respectively. At the receiver, a single optical modulation analyzer (OMA) acquires the mode channels using a polarization/time interleaver[25,38] when all four TE data modes are transmitted (Methods and Supplementary Fig. 6a). The data channels are then jointly recovered by multiple-input multiple-output (MIMO) equalization to compensate for crosstalk between TE modes[25,30]. Receiver digital signal processing (DSP) excludes frequency-offset estimation (FOE) and carrier-phase recovery (CPR), as detailed in Methods.

We first demonstrate self-homodyne transmission over a millimetre-scale multimode waveguide. The 1.8-μm-wide multimode waveguide has a total length of 1.46 mm and incorporates ten inverse-designed 90° bends (Fig. 3b), each with an effective radius of 20 μm (Methods and Supplementary Note 1). Signals at 46 Gbaud are launched into all four TE data modes at five carrier frequencies between 191.15 and 195.40 THz, spanning approximately 34 nm, and jointly recovered using 4 × 4 MIMO equalization. With 32-QAM or 64-QAM, all 20 mode–frequency combinations achieve normalized generalized mutual information (NGMI)[39] above the threshold of 0.8892 for concatenated soft- and hard-decision forward error correction (FEC)[40] (Fig. 3c), confirming that compact routing is compatible with broadband multimode transmission.

To assess the role of LO purity in this transmission scheme, we compare a fibre-delivered LO that bypasses the chip with LOs transmitted through the on-chip $TM_0$, $TE_0$ and $TE_3$ channels of a back-to-back DP-mMUX chip. The data channels remain fixed at $TE_1$ and $TE_2$, carrying 80-Gbaud QPSK, 16-QAM and 32-QAM signals, with an identical 2 × 2 MIMO equalizer configuration and fixed signal and LO powers at the OMA inputs across all LO paths. At the operating wavelength of 1542.14 nm, the measured transmission matrix shows approximately 15–22 dB lower leakage from the $TE_1$ and $TE_2$ data inputs into the $TM_0$ output than into the $TE_0$ and $TE_3$ LO outputs (Fig. 3d). The $TM_0$ LO achieves NGMI close to the fibre baseline across all tested modulation formats, with a mean NGMI only 0.013 lower for 32-QAM (Fig. 3e). By comparison, both TE LO configurations exceed the FEC threshold only for QPSK and exhibit pronounced NGMI degradation for 32-QAM. This degradation is accompanied by a 6–7 dB increase in residual power relative to the $TM_0$ LO (Fig. 3f) and blurred recovered constellations (Fig. 3g), consistent with residual SSBI from LO contamination that linear equalization cannot remove (Supplementary Note 2).

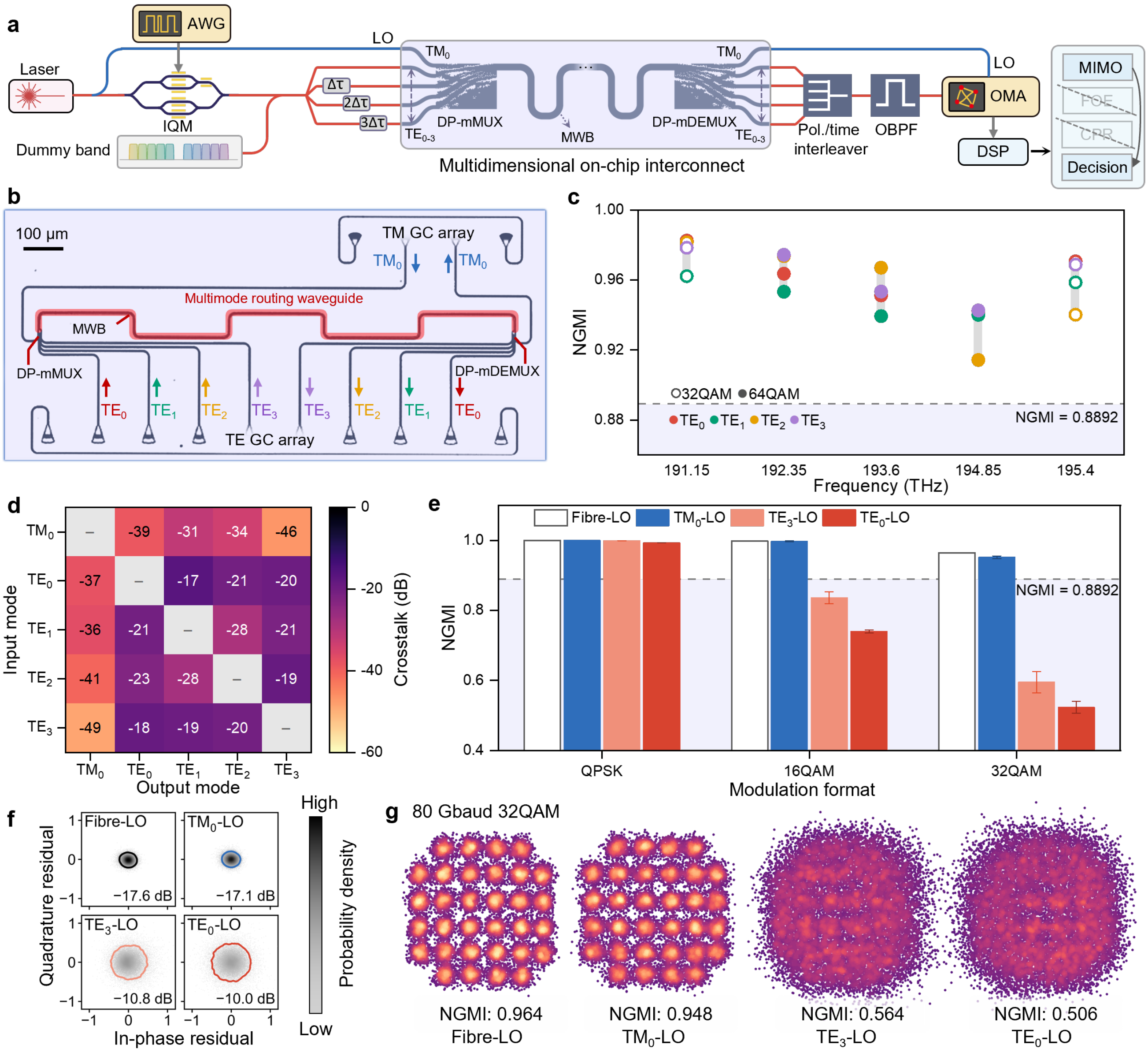

**Fig. 3 Self-homodyne multimode transmission with polarization-division LO routing. a,** Schematic of the self-homodyne multimode transmission experiment. Four decorrelated signal replicas and the LO are derived from the same laser and launched into the $TE_0$–$TE_3$ and $TM_0$ inputs of the DP-mMUX, respectively. The demultiplexed TE data modes are combined using a polarization/time interleaver and detected by a single OMA (Supplementary Fig. 6). AWG, arbitrary waveform generator; MWB, multimode waveguide bend; DP-mDEMUX, dual-polarization mode demultiplexer; OBPF, optical band-pass filter; OMA, optical modulation analyzer. **b,** Optical micrograph of the on-chip multimode routing circuit, comprising a DP-mMUX and DP-mDEMUX connected by a multimode waveguide incorporating ten inverse-designed 90° MWBs. **c,** Measured NGMI for self-homodyne transmission of four TE data modes through the multimode routing circuit at five carrier frequencies. Open and filled circles denote 32-QAM and 64-QAM signals, respectively; grey bars span the four modes. **d,** Measured crosstalk matrix of a packaged back-to-back DP-mMUX chip at 1542.14 nm. **e,** Comparison of measured NGMI under four LO configurations: a fibre-delivered LO and LOs transmitted through the on-chip $TM_0$, $TE_3$ and $TE_0$ modes. Bars and error bars indicate the mean and range across the two data modes, respectively. **f,** Distributions of the residual error for 32-QAM signals under the four LO configurations, each panel including data from both $TE_1$ and $TE_2$. The residual is defined as the error remaining after the linear response to the transmitted signals is removed from each received symbol. Its in-phase and quadrature components are normalized to the r.m.s. signal amplitude. Contours enclose 90% of the samples; annotated values indicate the mean residual power relative to the signal power. **g,** Recovered 80-Gbaud 32-QAM constellations for the worst-performing data mode ($TE_1$) under the four LO configurations.

## Terabit-per-second single-wavelength transmission

We next demonstrate high-speed coherent-lite multimode transmission with the on-chip $TM_0$ LO through the back-to-back DP-mMUX chip. Using a 4 × 4 MIMO equalizer operating at one sample per symbol (Methods), we quantify the tap count required for all four TE data modes to exceed the NGMI threshold of 0.8892 (Fig. 4a). Terabit-per-second operation on a single wavelength is achieved with single-digit tap counts, demonstrating the feasibility of low-complexity equalization for multimode coherent transmission. Specifically, QPSK at 135 Gbaud and 16-QAM at 90 Gbaud achieve single-wavelength aggregate line rates of 1.08 and 1.44 Tb $s^{-1}$, respectively, with only nine taps. To compensate for intersymbol interference arising from the limited bandwidth of the transmitter and receiver front ends, we extend the equalizer to 57 taps, enabling 16-QAM at 120 Gbaud and 32-QAM at 100 Gbaud. Further increases in tap count yield only marginal NGMI gains, so we adopt 57 taps for the subsequent single-wavelength and wavelength-division multiplexing (WDM) experiments.

With this fixed equalizer length, we compare single-wavelength transmission using the on-chip $TM_0$ LO and a fibre-delivered LO under different modulation formats and symbol rates (Fig. 4b). Across the measured symbol rates, the four-mode mean NGMI differs by less than 0.017 between the on-chip and fibre-delivered LO configurations. Together with the comparable recovered constellations (Fig. 4b(i–iii)), this agreement indicates that on-chip LO delivery via PDR introduces a negligible transmission penalty even at high symbol rates. At 100-Gbaud 32-QAM, the estimated 4 × 4 intensity transfer matrices show that MIMO equalization suppresses mode mixing and produces a predominantly diagonal effective channel response (Fig. 4c). All four modes exceed the NGMI threshold, yielding a line rate of 500 Gb $s^{-1}$ per wavelength per spatial mode and an aggregate line rate of 2 Tb $s^{-1}$ on a single wavelength. Constellations for all four modes at selected symbol rates are provided in Supplementary Fig. 9. The residual frequency-offset measurement and DSP comparisons further support reception without dedicated FOE or CPR (Supplementary Fig. 10).

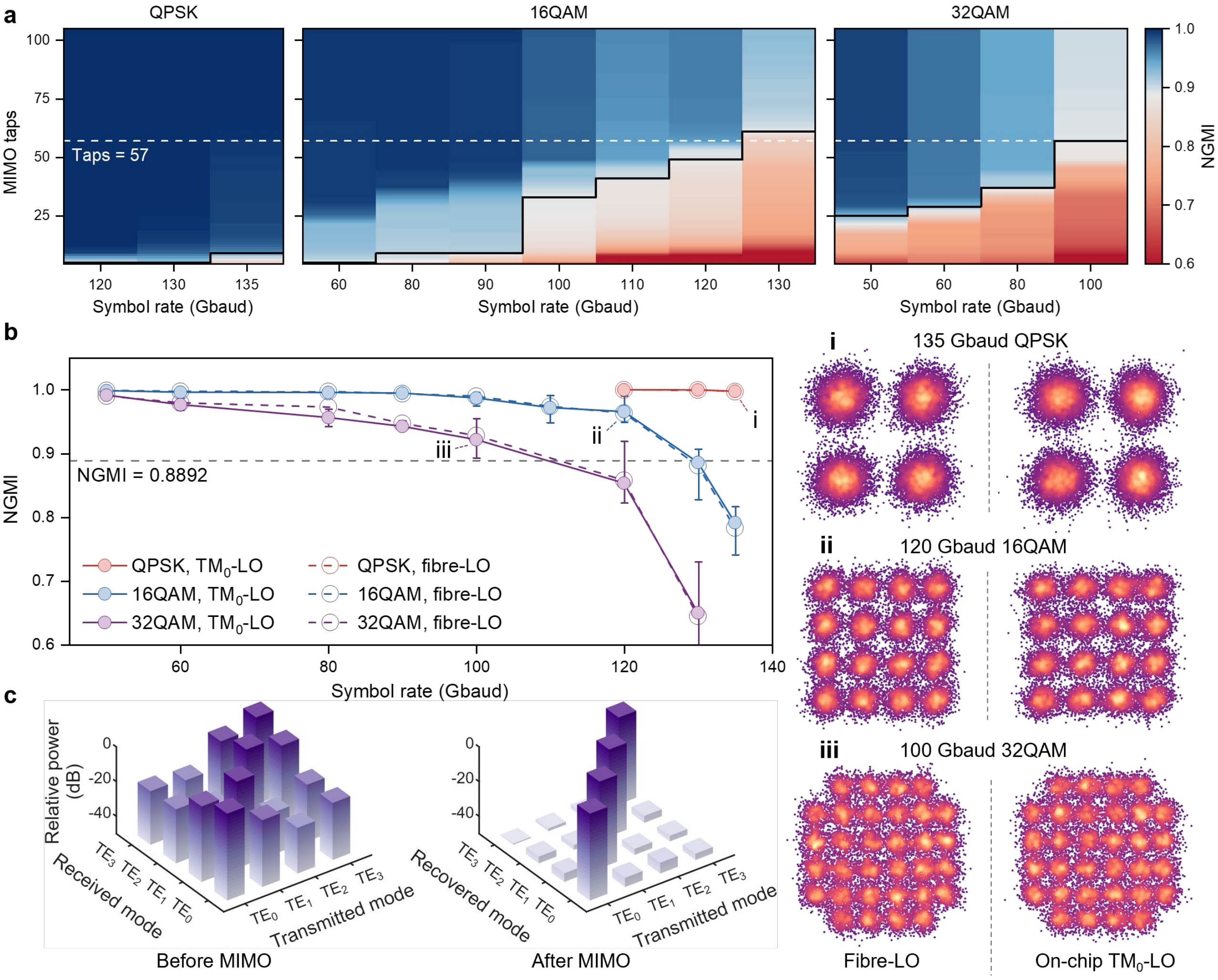

**Fig. 4 Experimental results for single-wavelength coherent transmission with the on-chip LO. a,** Worst-mode NGMI versus symbol rate and MIMO tap count for QPSK, 16-QAM and 32-QAM using the on-chip $TM_0$ LO. Each column corresponds to a measured symbol rate, displayed with equal widths. Black lines indicate the minimum tap counts required to reach the NGMI threshold of 0.8892; white dashed lines indicate the 57-tap setting. **b,** Comparison of NGMI versus symbol rate using the on-chip $TM_0$ LO and a fibre-delivered LO for QPSK, 16-QAM and 32-QAM. Symbols represent the mean NGMI across the four TE data modes; error bars span the minimum and maximum NGMI values for the on-chip $TM_0$ LO. Insets show recovered constellations of the worst-performing data mode for **(i)** 135-Gbaud QPSK, **(ii)** 120-Gbaud 16-QAM and **(iii)** 100-Gbaud 32-QAM, using the fibre-delivered LO (left) and on-chip $TM_0$ LO (right). **c,** Estimated 4 × 4 intermodal intensity transfer matrices before (left) and after (right) MIMO equalization for 100-Gbaud 32-QAM using the on-chip $TM_0$ LO, normalized to the target-mode power in each row.

## Hundred-terabit multidimensional transmission

Finally, we scale the aggregate capacity by combining high-order self-homodyne transmission with dense wavelength-division multiplexing (DWDM) and MDM. This requires maintaining LO purity and multimode transmission performance across the full C band under adjacent-channel interference and full spectral loading. The measurements cover 100 wavelengths on a 50-GHz grid spanning 191.15–196.10 THz, with 46-Gbaud signals transmitted across four TE data modes and the co-propagating LO delivered through $TM_0$ using PDR. For each measurement, the optical carrier of the channel under test (CUT) is modulated by one IQ modulator, while the two adjacent carriers are modulated by another. The remaining wavelength slots are filled with spectrally shaped light from an amplified spontaneous emission (ASE) source to emulate full C-band loading (Fig. 5a). All 400 wavelength-mode channels are characterized by scanning the CUT across the band under this loading condition. As the CUT is scanned across the band, the two modulated neighbouring carriers are tuned with it, and the ASE spectrum is reshaped to fill the remaining wavelength slots.

High-order self-homodyne transmission is achieved across all 100 wavelengths using the same 57-tap $4 \times 4$ linear MIMO configuration, without FOE or CPR. At each wavelength, the modulation format is selected to increase the aggregate capacity while ensuring that all four TE data modes exceed the NGMI threshold. The resulting allocation supports 64-QAM at 65 wavelengths and 32-QAM at the remaining 35, with all 400 wavelength-mode channels achieving NGMI above 0.8892 (Fig. 5b). Recovered constellations for wavelength channels 1, 50 and 100 show four-mode signal recovery at the band edges and centre (Fig. 5c). Across the four TE data modes, 64-QAM and 32-QAM provide aggregate line rates of 1.104 and 0.920 Tb $s^{-1}$ per wavelength, respectively. Summing across the 100 wavelengths yields an aggregate line rate of 103.96 Tb $s^{-1}$ within a single silicon waveguide, corresponding to an aggregate line-rate spectral efficiency of 20.79 bit $s^{-1}$ $Hz^{-1}$ over the 5-THz WDM allocation. Accounting for FEC overhead[40] at a code rate of 0.8428 gives an aggregate net data rate of 87.62 Tb $s^{-1}$.

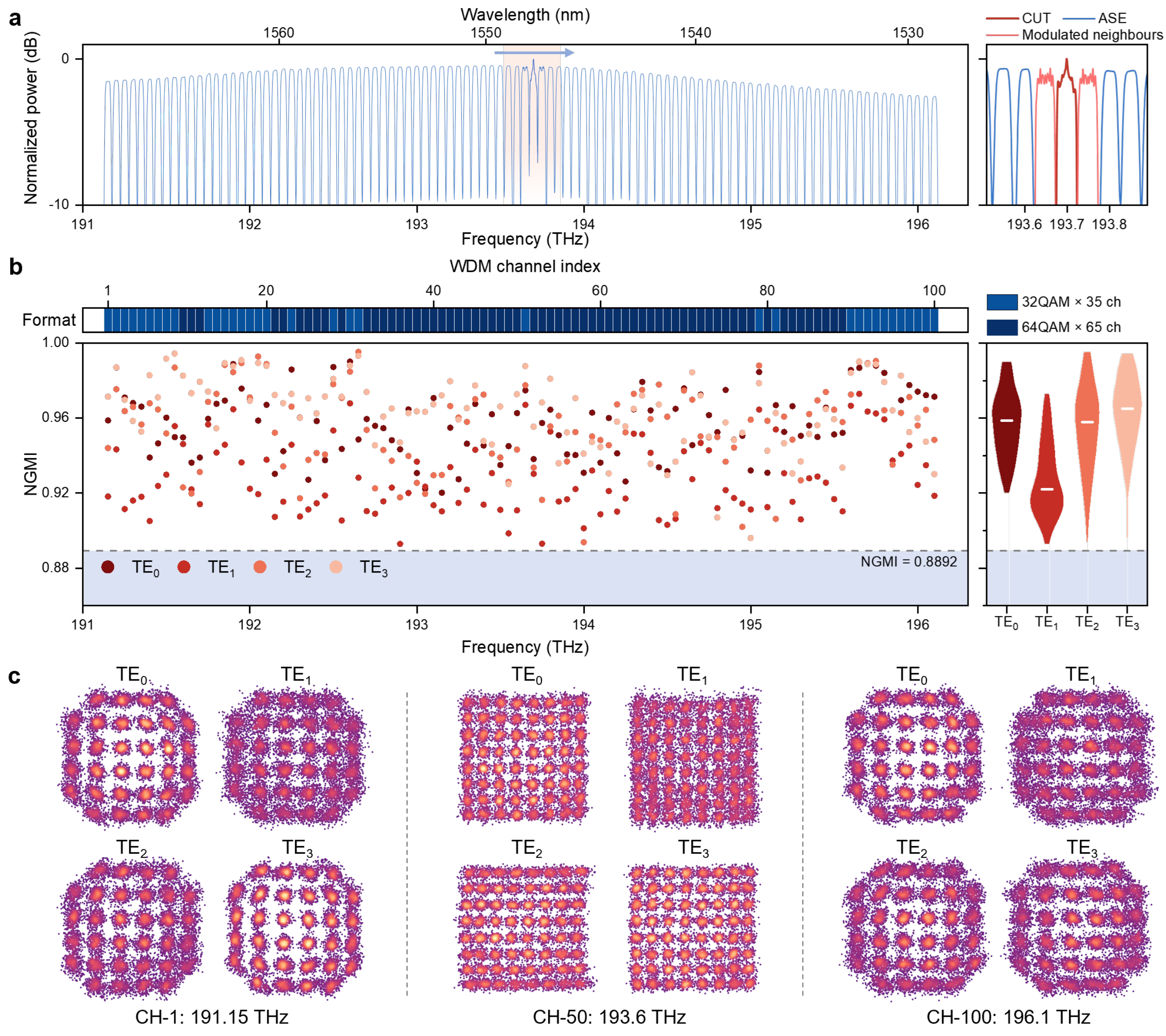

**Fig. 5 Experimental results for hundred-terabit multidimensional self-homodyne transmission. a,** Normalized optical spectrum of the 100-channel DWDM signal at the transmitter, spanning 191.15–196.10 THz on a 50-GHz grid. Right, enlarged view of the shaded region. The channel under test (CUT) is generated by one IQ modulator and its two adjacent channels by another, while the remaining wavelength slots are filled with spectrally shaped ASE to emulate full C-band loading. **b,** Measured NGMI of 46-Gbaud signals in the four TE data modes across 100 wavelength channels. The top bar shows the wavelength-dependent modulation-format allocation, while the dashed line marks the NGMI threshold of 0.8892. Right, NGMI distributions for each mode across the 100 channels, with white lines indicating the medians. **c,** Recovered constellations of the four TE data modes for channels 1 (32-QAM), 50 (64-QAM) and 100 (32-QAM).

## Discussion

As further scaling of per-lane symbol rate becomes increasingly constrained, optical interconnect capacity can be increased along two complementary dimensions: greater parallelism or more information carried per symbol. Figure 6 compares representative scale-in and scale-up silicon photonic interconnects in terms of net capacity per waveguide/fibre and average net bits per symbol (Supplementary Note 4). Commercial IM-DD optical I/O based on WDM-only non-return-to-zero (NRZ) transmission is limited to wavelength parallelism and one bit per symbol[41,42]. Research demonstrations have extended capacity through additional wavelength[43] and spatial dimensions[27], reaching 38.2 Tb $s^{-1}$ per waveguide with WDM–MDM[26]. Yet the benchmarked IM-DD implementations remain below 2 net bits per symbol, leaving limited headroom for further per-lane scaling and constraining aggregate capacity growth. Coherent modulation increases the information carried by each symbol and thereby relaxes this constraint. In the scale-up regime, coherent MDM over few-mode fibre has demonstrated terabit-per-second transmission with substantially higher net bits per symbol[44]. At the chip scale, integrated coherent MDM using 16-QAM has achieved net capacities of 2.162 and 1.23 Tb $s^{-1}$ with off-chip LO delivery[25,30], which requires a separate fibre path or an independent laser for each link and is therefore difficult to scale in densely integrated interconnects. Addressing this gap, our coherent-lite multidimensional architecture combines high-order coherent modulation with large-scale multidimensional parallelism in the same on-chip optical path, achieving 87.62 Tb $s^{-1}$ net capacity at an average of 4.76 net bits per symbol across 400 wavelength-mode data lanes within a single silicon waveguide.

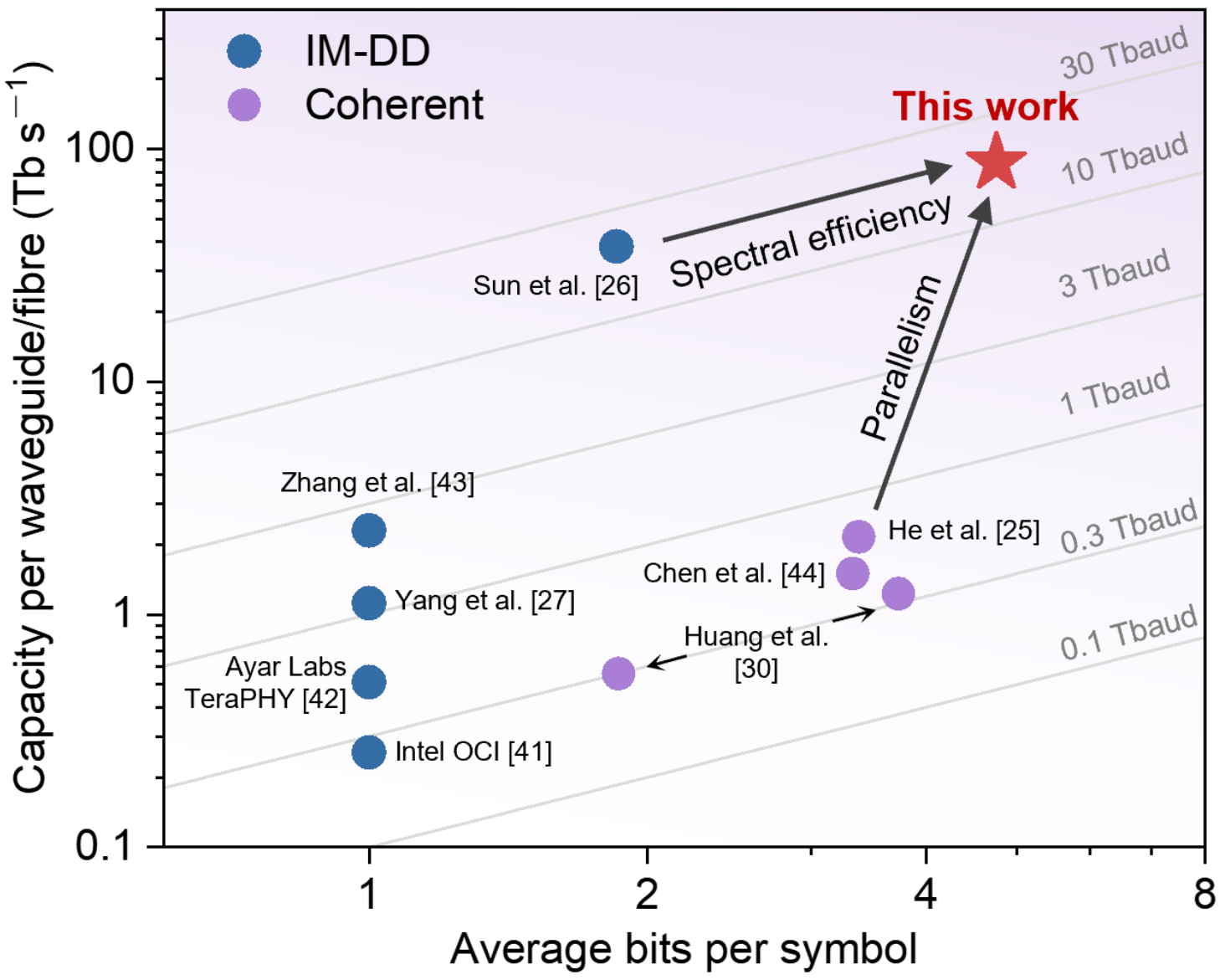

**Fig. 6 Comparison of scale-in and scale-up silicon photonic interconnects.** Capacity per waveguide/fibre versus average bits per symbol for optical I/O products and research demonstrations[25–27,30,41–44]. Both metrics are reported as net values accounting for FEC overhead. Grey diagonal lines indicate constant aggregate symbol rates, used here as a measure of transmission parallelism. This work scales capacity through both parallelism and spectral efficiency.

Extending coherent-lite multidimensional interconnects from on-chip links to in-package connectivity across chiplets and photonic interposers will require advances in routing reach, signal processing and per-lane throughput. As routing distances within the package approach the centimetre scale, low-loss multimode transport of both data and LO becomes increasingly important. Passive photonic platforms such as silicon nitride (SiN) are well suited to this role owing to their low propagation loss and suitability for long passive routes[45,46]. Beyond on-chip routing, the carrier-recovery-free operation enabled by self-homodyne detection creates a natural opportunity for optical-domain MIMO processing under the stringent energy-per-bit and latency requirements of in-package interconnects. Reconfigurable interferometric meshes have already demonstrated optical spatial-mode unscrambling[44,47], while recent optical signal processors have demonstrated DSP-free real-time photonic equalization of temporal impairments at terabit-per-second throughput[48]. Integrating such optical-domain processing with coherent-lite reception could therefore reduce the energy and latency of multimode signal recovery. In parallel, continued advances in electro-optic and optoelectronic front ends can further increase per-lane throughput[49]. Integrated IQ modulators with bandwidths exceeding 100 GHz and high-speed coherent receiver front ends are extending coherent operation beyond 100 Gbaud[50–53]. Together, these developments point towards coherent-lite interconnects with longer routing reach, lower processing overhead and higher per-lane capacity, further bringing coherent optics closer to compute.

# Methods

### Device design and simulations

The DP-mMUX was optimized using EG-ADO[26], comprising topology optimization, edge-guided conversion and digital optimization (Supplementary Fig. 1). The figure of merit (FOM) combined the conversion efficiencies from the five input waveguides to their designated $TE_0$–$TE_3$ and $TM_0$ modes, obtained by projecting the output fields onto the corresponding eigenmodes at a common transverse monitoring plane and averaged over 11 wavelengths across 1540–1560 nm. Forward and adjoint fields were calculated using 3D-FDTD simulations in ANSYS Lumerical on a 20-nm in-plane grid, and the resulting gradients were used to update the permittivity distribution with an L-BFGS optimizer, while a progressively sharpened Heaviside projection drove the distribution towards the silicon and $SiO_2$ states. The continuous design was then mapped onto 100-nm digital pixels guided by Canny edge detection, and the pixels near material boundaries were determined by the direct binary search (DBS) method. In addition to the five-channel DP-mMUX used in this work, a seven-channel device multiplexing $TM_0$ and $TE_0$–$TE_5$ was designed using the same procedure (Supplementary Fig. 4). The multimode $90^\circ$ bend, with an effective radius of 20 μm, was optimized using particle swarm optimization (PSO) by varying the waveguide widths at nine positions along a quasi-uniform B-spline centreline (Supplementary Fig. 5).

### Device fabrication and optical characterization

The devices were fabricated at Tianjin H-chip Technology Group on a silicon-on-insulator (SOI) wafer with a 220-nm silicon device layer and a 2-μm buried oxide layer. Two aligned electron-beam lithography (EBL) steps were used to define the photonic structures. In both steps, a 250-nm-thick ZEP resist was exposed with a beam current of 2 nA. The grating couplers were formed using a 70-nm shallow etch, while the remaining silicon structures were fully etched through the 220-nm device layer by inductively coupled plasma (ICP) etching. After resist removal and scanning electron microscopy inspection, an 800-nm $SiO_2$ upper cladding was deposited by plasma-enhanced chemical vapour deposition.

The DP-mMUX was characterized in a back-to-back MUX–DEMUX circuit using a six-axis fibre-coupling stage. On-chip TE and TM grating couplers were used at the corresponding channel ports, both with a coupling loss of approximately 6 dB per facet at 1550 nm. Broadband transmission spectra were measured using a spectrally flat ASE source (OVLINK ASE-CL-PM). For each input port, a polarization controller was used to align the incident polarization with that required by the corresponding grating coupler, and the spectrum at each output port was recorded using an optical spectrum analyzer (Yokogawa AQ6370D). By sequentially exciting each input port and measuring all output ports, the broadband transmission matrix was obtained. To account for the different coupling losses of the TE and TM grating couplers, separate reference waveguides incorporating a pair of TE grating couplers or a pair of TM grating couplers were measured under identical conditions. TE–TE and TM–TM transmissions were normalized to their respective reference spectra. For TE–TM and TM–TE transmissions, which each involve one TE and one TM grating coupler, the normalization reference was taken as the arithmetic mean of the TE–TE and TM–TM reference spectra in decibels. The normalized $TM_0$ transmission fluctuates slightly around 0 dB due to variations in grating-coupler efficiency.

**Detailed experimental setup**

The detailed experimental setup for self-homodyne multidimensional transmission is shown in Supplementary Fig. 6. Equipment models are listed in Supplementary Table 2. For four-mode measurements, the TE-mode outputs were polarization-multiplexed in pairs and interleaved into two time slots, with two modes acquired simultaneously in each slot by a single OMA (Keysight N4391B). Light from a tunable laser (ID Photonics, CoBrite MX) with an output power of 16 dBm was split into signal and LO branches. In the signal branch, an arbitrary waveform generator (AWG, Keysight M8199A) with a sampling rate of 256 GSa $s^{-1}$ drove a single-polarization IQ modulator to generate the CUT. The modulated signal was amplified using an erbium-doped fibre amplifier (EDFA) and gated by an acousto-optic modulator (AOM, OVLINK SGTF-100) using a 200-kHz rectangular pulse train with a 48% duty cycle, generated by a function generator (RIGOL DG4202).

For both single-wavelength and WDM measurements, the output of an ASE source (OVLINK ASE-CL-PM) was combined with the gated signal to suppress transient power excursions associated with AOM gating. For WDM measurements, two additional carriers at ±50 GHz from the CUT were jointly modulated by a second IQ modulator to generate the neighbouring channels, while the remaining wavelength slots were filled with ASE shaped by a wavelength-selective switch (Finisar WaveShaper 4000A) on the same 50-GHz grid. The combined signal was amplified using an EDFA and split into four replicas, decorrelated by relative fibre delays of 0, 25, 50 and 75 ns, and polarization-aligned before being launched into the $TE_0$–$TE_3$ inputs of the DP-mMUX. At the receiver, $TE_0$/$TE_1$ and $TE_2$/$TE_3$ were polarization-multiplexed in pairs using two polarization beam combiners. A 500-m fibre delay introduced a relative delay of approximately 2.5 μs between the two mode pairs, which were subsequently recombined into two temporal slots. With the AOM-generated optical bursts being 2.4 μs long, a short guard interval remained within the 2.5-μs spacing between the two time slots, enabling clear frame separation. The recombined signal was amplified by a receiver-side EDFA and then filtered using an optical band-pass filter (EXFO XTM-50) before being fed to the signal input of the OMA.

In the LO branch, an identical AOM gated the LO synchronously with the signal branch using the same pulse train from the function generator. The gated LO was then split into two paths, with a 500-m fibre delay introduced in one path to match the receiver-side time interleaving. The polarization states of the two replicas were independently adjusted before recombination. The recombined LO was passed through a polarizer to align its polarization with the principal axis of the subsequent polarization-maintaining EDFA (PM-EDFA), amplified, and then polarization-adjusted to TM before being launched into the $TM_0$ input of the DP-mMUX. The LO co-propagated with the four TE data modes through the multimode waveguide. The total optical powers before fibre-to-chip coupling were approximately 12 dBm in the LO branch and 4 dBm in each TE data branch. After demultiplexing, the recovered $TM_0$ LO was amplified using another PM-EDFA and supplied to the external LO input of the OMA. The resulting electrical waveforms were digitized using the UXR0594A oscilloscope of the OMA at 256 GSa $s^{-1}$ with a 59-GHz analog bandwidth and processed offline.

For the LO-purity comparison (Supplementary Fig. 6b), only the $TE_1$ and $TE_2$ data channels were used and acquired simultaneously on the two polarization tributaries of the OMA, without time interleaving. The LO was delivered through a fibre bypass or transmitted through the on-chip $TM_0$, $TE_0$ or $TE_3$ channel.

**Digital signal processing**

The transmitter and receiver DSP flows are summarized in Supplementary Fig. 7. At the transmitter, pseudo-random binary sequence (PRBS) data were generated and mapped onto the target QAM constellation. The resulting symbols were pulse-shaped using a root-raised-cosine (RRC) filter with a roll-off factor of 0.04 and subsequently resampled to match the sampling rate of the AWG before electrical waveform generation. At the receiver, the captured I/Q waveforms were processed offline. For the four-mode measurements, the two time slots were extracted to obtain the four tributaries, which were resampled, corrected for I/Q imbalance and passed through an RRC matched filter. The frame starts were then located by correlation, and the receiver-side interleaving delay was compensated while preserving the transmitter-side decorrelation delays (Supplementary Fig. 8). A common one-frame window was jointly processed at one sample per symbol using a linear MIMO equalizer trained by least squares (LS) to compensate for inter-mode crosstalk and residual linear distortions.

The recovered symbols were used to evaluate the transmission performance in terms of NGMI. NGMI was estimated from the equalized symbols using bit-wise log-likelihood ratios computed with the max-log approximation under a Gaussian noise assumption, with the noise variance estimated from the symbol-decision residuals.

## Data Availability

The data that support the findings of this study are available from the corresponding author on request. Source data are provided with this paper.

## Code Availability

The open-source code for the inverse design algorithm is available via GitHub at https://github.com/BryvinSal/EGADO. Additional code used in this work is available from the authors upon reasonable request.

## Acknowledgements

This work was supported by the National Natural Science Foundation of China (Grant Nos. 62625114, 62525102 and 625B2056) and the Guangdong Major Project of Basic and Applied Basic Research (2025B0303000008).

## Author contributions

Y.H. and A.S. contributed equally to this work.
J.W.Z., A.S. and Y.H. conceived the study. A.S., J.H. and Y.Y. designed and simulated the photonic devices. A.S. and B.D. performed the theoretical analysis. A.S., J.H. and Y.H. characterized the devices. Y.H., H.Q., A.Y., A.S., P.S. and H.C. performed the transmission experiments. Y.H. and A.S. performed the digital signal processing and data analysis. J.H.Z., C.S., Z.L., C.L. and X.Z. provided advice on the experiments and data analysis. Z.H., N.C., S.Y. and J.W.Z. supervised the project. All authors discussed the results and contributed to the writing of the manuscript.

## Competing interests

The authors declare no competing interests.